\documentclass[fleqn,twoside]{article}
\usepackage{espcrc2}

\usepackage{graphicx}
\usepackage[figuresright]{rotating}

\newcommand{\AmS}{{\protect\the\textfont2
  A\kern-.1667em\lower.5ex\hbox{M}\kern-.125emS}}

\title{Study of CMS sensitivity to neutrinoless $\tau$ decay at LHC}

\author{R.Santinelli \address{Physics Department, University of Perugia and INFN Perugia, \\
        1 via Pascoli, 06100 Perugia, Italy}}
       
\begin{document}

\begin{abstract}
The Large Hadron Collider (LHC), scheduled to start operation in 2006, is foreseen to provide in the first year of running a total of $\sim 10^{12}$ $\tau$ leptons.\\
CMS (Compact Muon Solenoid) is a general-purpose experiment designed to study proton-proton and heavy-ion collisions at LHC.
Even if the Susy particles and Higgs searches togheter with the B-physics present its main goal, the large amount of $\tau$-lepton, could allow a systematic study of tau-physics.
We have performed a full simulation of CMS using GEANT 3 package and the object-oriented reconstruction program ORCA to study the sensitivity to neutrinoless tau decay $\tau \rightarrow \mu^+ \mu^- \mu^- $ and $\tau \rightarrow \mu \gamma$.
We present the analysis developed for these channels and the results obtained.
 
\vspace{1pc}
\end{abstract}

\maketitle

\section{THEORETICAL INRODUCTION}
In the Standard Model the neutrinoless decay of tau lepton are not foreseen because they would violate the {\it Charged Lepton Flavour} and/or the {\it Baryonic Number}.
Until now there is no experimental evidence for such processes.
The Noether theorem states that for every conservation law there must an associated symmetry and conversely. In the Standard Model we do not have a symmetry associate to the Charged Lepton Flavor Conservation law.
It is simply a built in characteristics of the theory.
Beyond the SM a large number of existing theories, can explain and accomodate them.
An uncomplete list will include the Standard Model with right handed neutrinos, the Fourth generation neutrino theory, See-Saw type II models, GUT-theory, Susy with R-parity broken, Super Strings theories, Leptoquarks, Technicolor.\\
Every model in the above list provides or too an optimistic or too a pessimistic expectation of the neutrinoless decay Branching Ratios.
Nevertheless, under the experimental point of view, the mSUGRA scenario with right handed neutrinos offers the closest forecast to the future experimental sensitivity.
In this theory the vertices causing the violation of charged lepton flavour are due to non diagonal terms in the sleptons matrix mass which cannot be diagonalized simultaneously with the mass matrix of neutrinos. Mixing vertices arise at 1-loop radiative corrections from Yukawa couplings between sleptons and neutrinos.\cite{Ellis}  
With an opportune choice of the input universal parameters and with a texture describing the mixing in neutrino sector in accordance with the Super Kamiokande experiment, it possible to find a value of the Branching Ratio for the $\tau \rightarrow \mu \gamma$ around $10^{-7}$ while a 200 times lower value is expected for the $\tau \rightarrow \mu^+ \mu^- \mu^- $.

\section{CMS}

The main features of this detector are the strong (4 Tesla) magnetic field, ensuring high momentum resolution for charged track, the efficient muon system, providing a very good reconstruction of muon and the very small global size.\\
The detector consists of a silicon tracker with an embedded pixel detector, a crystal PbWO$_4$ electromagnetic calorimeter (ECAL) and a copper-scintillator hadron calorimeter(HCAL). A sophisticated four station muon system made of tracking chambers, the drift tube (DT) and the cathode strip chamber (CSC) and of dedicated trigger chamber (the resistive plate chamber or RPC), is located outside the solenoidal magnetic field. 
The overall dimensions of the CMS detector are: 21.6 m in length, 14,6 m in diameter with a total weight of 14,500 tons.
Both the analysis I will present here will rely heavily on the performances of the muons subdetector for its high efficiency to detect isolated muons (95\%) and its low probability to misindentify pions or kaons as muons.

Another very important component of CMS detector is its tracking system.
The performances of the CMS tracker play a fundamental role in the 
improvement of the Branching Ratio sensitivity to tau in three muons decay channel.

The main requirements are:
\begin{itemize}
\item high radiation resistance because the tracker will be the part of CMS closest to the primary interaction point and it will be therefore operating in a very high radiation enviroment
\item a low material budget, to minimize the photon conversion and bremsstrahlung. 
\item high momentum resolution
\item high secondary vertex reconstruction efficency
\end{itemize}

The expected resolution on the transverse momentum of muons will be $\frac{\sigma_{Pt}}{Pt}=$1.5\% for 10 GeV muons and the resolution on the secondary vertex reconstruction about 200 $\mu m$.
The particle detection efficiency is expected to be 95\% for isolated charged tracks and 90\% for tracks inside a jet.\\

As for what regards the calorimeters, they become of the ut most importance in 
the $\tau \rightarrow \mu \gamma$ analysis, where we require a 
good photon reconstruction.\\
The transverse energy resolution for electron/photon could be 
parameterized as:
\begin{equation}
\frac{\sigma^{ECAL}_E}{E}=\frac a {\sqrt E} \otimes b \otimes \frac c {E}
\end{equation}
The ``stochastic term''$a$ arises from photoelectron statistics and shower fluctuaction. The ``constant term'' $b$ has contributions from non-uniformities and from shower leakage. The ``noise term'' $c$ is due to electronics noise and pile-up.
The design goal for the CMS ECAL barrel and endcaps are $a=2.7\%$ and $a=5.7\%$ respectively and $b<0.55\%$ for both section of the detector. Expressing the noise as trasverse energy, the goals for the {\it c} term are at low luminosity $c$=155MeV for the barrel part and 205MeV for the end-caps. 
The phi-angle resolution is expected to be 1.3 mrad.

\section{$\tau \rightarrow \mu^+ \mu^- \mu^- $}

In Table 1 we summarize the main sources of tau leptons at LHC as we 
have found using PYTHIA 6.152 generator and the branching ratio 
listed in the PDG tables.

\begin{table*}[bht]\label{tab:1}
    \caption{Main sources of $\tau$-leptons at LHC}
    \begin{center}
    \begin{tabular}{c c c c c c } \hline 
     Meson (M)           & D$_S$ & D$^+$ & B$^0$ & B$_S$ & B$^+$ \\ \hline
     BR($M \rightarrow \tau +X$) & 7.0\%    &   0.2\%   & 2.7\%   & 1.5\%     & 2.7\% \\ \hline
 $\sigma(M \rightarrow \tau +X) / \sigma(pp \rightarrow \tau +X)$& 77\%    &   3\%   & 9\%   & 2\%     & 9\% \\ \hline 
      \end{tabular}
    \end{center}
\end{table*}
The total cross section will be of around 120 $\mu b$ for a total 
amount of tau produced after one year of low luminosity of LHC 
(integrated luminosity of 10 fb$^{-1}$) 
of 10$^{12}$ tau lepton.
Because we will only be capable to trigger on the higt p$_t$ particles, we have concentrated on the following signal sources:
\begin{itemize}
\item W source, $\sigma (pp \rightarrow W \rightarrow \tau \nu_{\tau} )$=19 nb \cite{nuclear}\\
\item Z source, $\sigma (pp \rightarrow Z \rightarrow \tau \tau )$=3nb
\item B source, $\sigma (pp \rightarrow B \rightarrow \tau D \nu$=24$\mu$b
\end{itemize}

The W source, as we shall see, represents the best source for the signal.

\subsection{The background}
The presence of three well reconstructed, collimated and isolated muons offers a very clear signature for the signal. 
According to the  results of \cite{muons} the main sources of $\mu$ that could contribute to the background will be:

\begin{itemize}
 \item heavy quarks mesons (D \& B) 
 \item pile-up effects from primary interaction
 \item Gauge bosons and Drell Yan (Z,W)
 \item $\tau$ production
 \item cosmic rays
 \item Susy Particles
\item Higgs decay
 \item non prompt muons (punchthrough pions non interacting in the calorimeters)  
\end{itemize}

Pile-up effects from primary interaction and tau source of muons are not considered for their low 
probability to have a final three muons topology.
The cosmic rays will be removed by timing.
W and Z source will give negligible contribution due to their distinctive high mass.
Susy particles and Higgs sources present a too low total production cross section
to be really dangerous at the present sensitivity attainable.\\
We focused then on heavy quarks mesons and on unphysical background. Events were simulated using the PYTHIA \cite{PYTHIA} kinematics generation followed by the GEANT3 (\cite{GEANT}) based simulation program CMSIM (\cite{CMSIM}) and the reconstruction program ORCA (\cite{ORCA}).

\subsection{Heavy quarks mesons events}
\begin{center}
\begin{figure}[h]
\includegraphics*[width=45mm,angle=270]{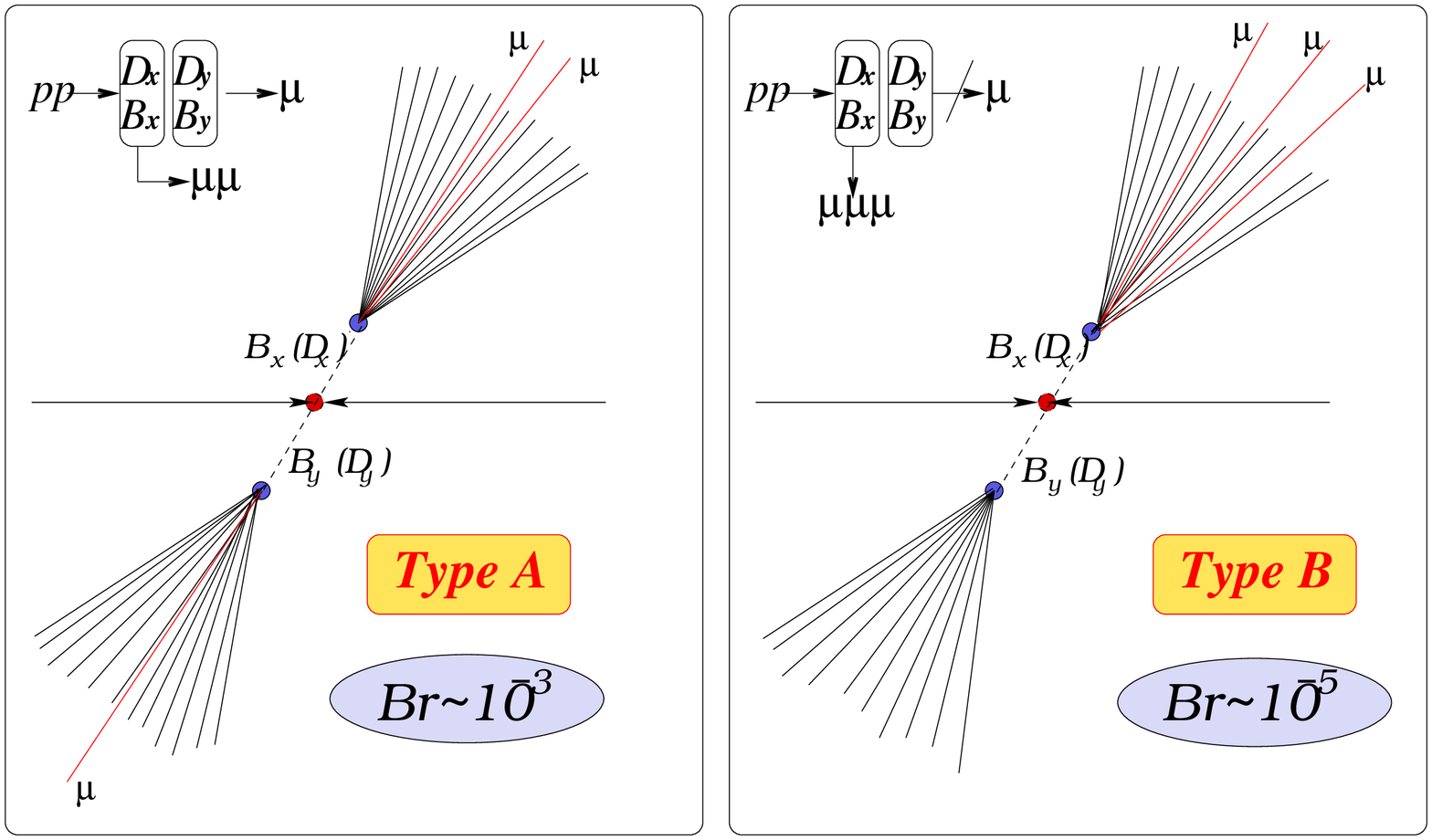}
\caption{Three muons from $b \overline b$ and $c \overline c$ events \label{fig:events}}
\end{figure}
\end{center}

We focus our investigations over events with only 3 muons in 
the final state.
This kind of events can occur in two different ways as depicted in fig.~\ref{fig:events}
 A preliminary study, conducted at kinematic generation level, has shown that an upper cut on the angle between two muons removes completly events of type A, leaving almost all the events of type B.
These can be generated by forcing decay cascades where some resonances  rare decay should be included like $\phi \rightarrow \mu \mu$
For the background we forced the following decays
\begin{itemize}
\item $pp \rightarrow D_X D_S \rightarrow D_X \mu \nu_{\mu} \phi \rightarrow D_X \mu \nu_{\mu} \mu \mu$
\item $pp \rightarrow B_X B_S \rightarrow B_X \mu \nu_{\mu} D_S \rightarrow B_X \mu \nu_{\mu} \mu \mu K$

\end{itemize} 
It is worthwhile to note that the choice of the $\phi$ as the resonance decaying into 2 $\mu$ is not unique, but one of the most important.
 We could use other resonances like $\omega, \eta, \eta^{\prime}$ but their contribution will be too small.
Moreover, to increase the final statistics, we required, at the generation level, three muons with a transverse momentum greater than 3 GeV. We will refer to these events as {\it preselected}.
Finally, we obtained from a fast simulation, the very important 
information that the $b \overline b$ events are negligibile if compared with $c \overline c$.
This is true because the reconstructed final three muons mass for $b \overline b$ events is shifted toward 4-5GeV values while the one from $c \overline c$ is peaked around 1.5 GeV.\\
We will consider then the events $pp \rightarrow D_X D_S \rightarrow D_X \mu \nu_{\mu} \phi \rightarrow D_X \mu \nu_{\mu} \mu \mu$ as the main source of background ({\it main background})

\subsection{Signal and analysis}
Different $\tau$-sources give rise to different signal signatures.
We have therefore adopted three different set of cuts appropriates for the different sources of signal. 

\subsubsection{$\tau$ from W}

\begin{center}
\begin{figure}[h]
\includegraphics[width=70mm,angle=0]{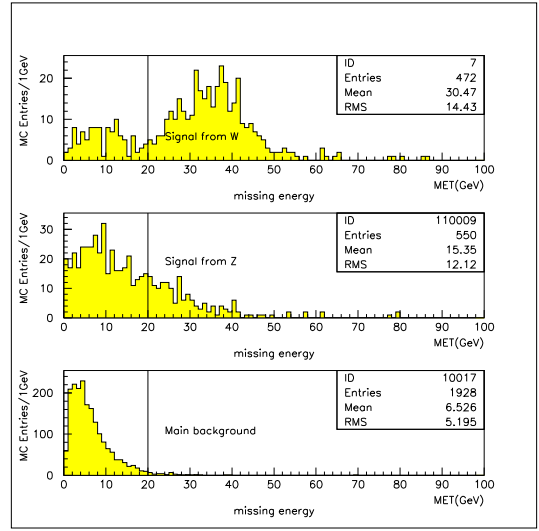}
\caption{Transverse missing energy \label{fig:missing}}
\end{figure}
\end{center}

These events are mainly characterized by an high value of missing energy.
Indeed the tau coming from W decay shares half of the initial boson's energy with the corresponding neutrino and this will be detected as an high missing energy, (see fig.~\ref{fig:missing})

Here we have for the signal from W the typical Jacobian peak around 35 GeV while the distribution is shifted toward low values for the main background and for the signal from Z.
The complete analysis implemented for these events require:
\begin{itemize}
\item trigger 
\begin{enumerate}
\item at least 3 muons passing the CMS trigger level 1 (L1)
\item 2 or more CMS level 3 muons (the final CMS trigger level for muons) (L3)
\end{enumerate}
\item identification
\begin{enumerate}
\item  only 3 well reconstructed and close muons candidates from tracker with a Pt$>$4GeV in the barrel region and Pt$>$2.5GeV in the endcaps
\item total charge equal $\pm$1
\end{enumerate}
\item topology
\begin{enumerate}
\item Common secondary vertex (not necessary to implement!)
\item no charged tracks inside a cone of $\Delta R= \sqrt{{\Delta \phi}^2+{\Delta \eta}}=0.4$ around the $\mu$ tracks.
($\phi$ is the azimuthal angle and $\eta$ is the pseudorapidity).
\end{enumerate}
\item missing transverse energy $>$20GeV
\item veto on phi-mass region (1020$\pm$25)MeV for any couple of muons
\item reconstructed three muons mass =1778$\pm$20MeV 
\end{itemize}
\begin{center}
\begin{figure}[h]
\includegraphics[width=70mm,angle=0]{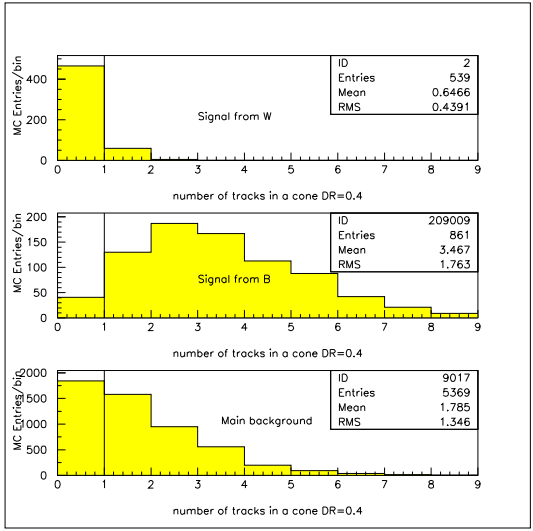}
\caption{Number of track inside a cone of opening =0.4 around the three muons \label{fig:iso}}
\end{figure}
\end{center}
\begin{center}
\begin{figure}[h]
\includegraphics[width=70mm,angle=0]{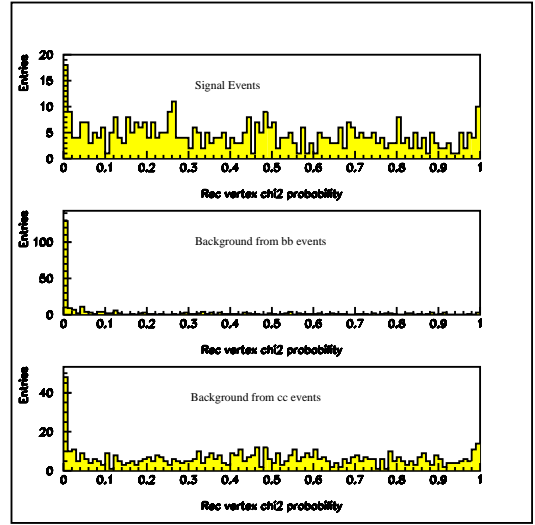}
\caption{Reduced chi squared probability of three muons to belong to same vertex \label{fig:chi}}
\end{figure}
\end{center}
\begin{center}
\begin{figure}[h]
\includegraphics[width=70mm,angle=0]{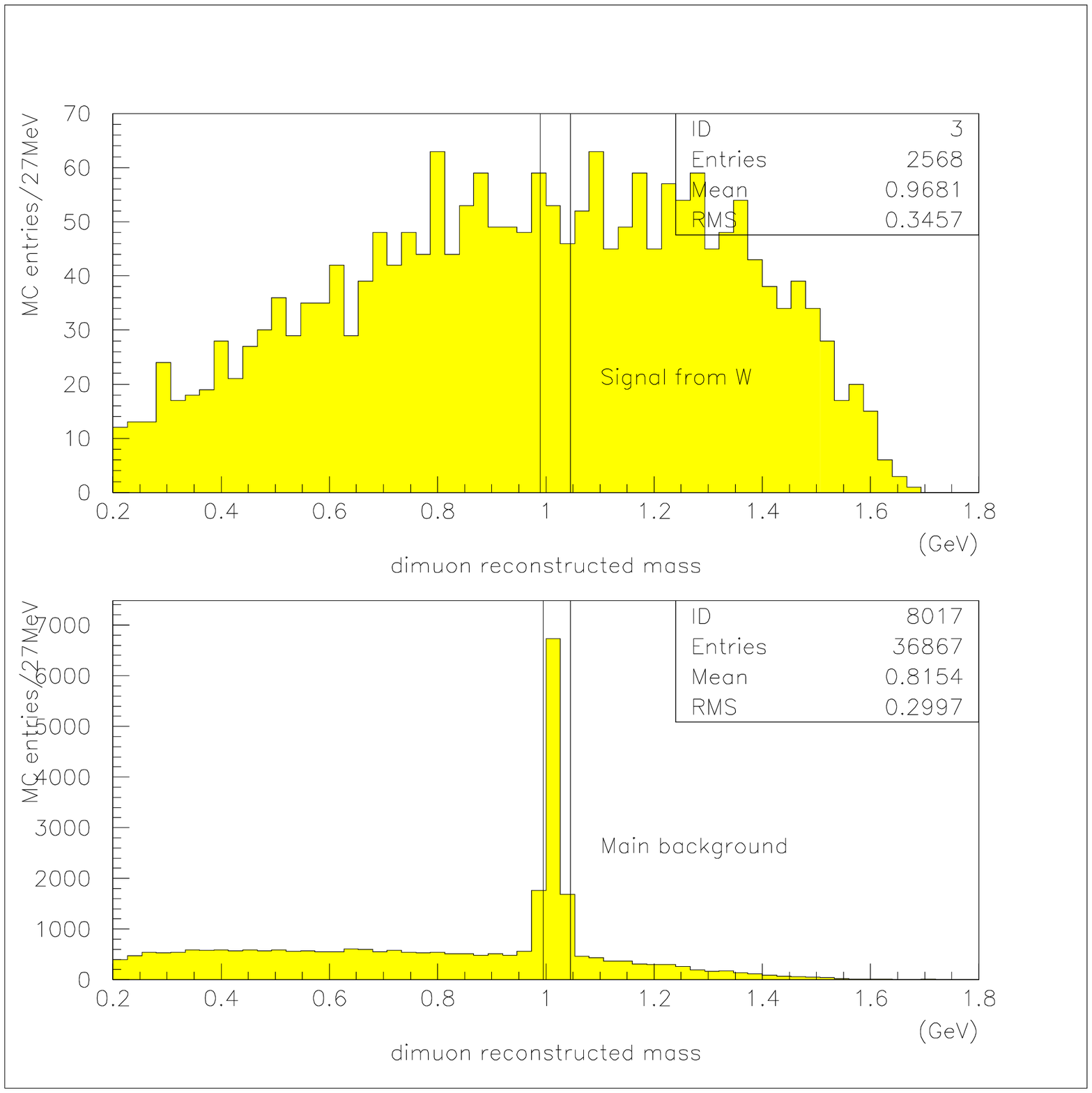}
\caption{Combinatorial dimuon mass \label{fig:massphi}}
\end{figure}
\end{center}

 The isolation cut is justified by the high multiplicity of background events and the low number of tracks for this kind of signal.
In figure~\ref{fig:iso} we compare the number of tracks around the three muons for the signal from W, for  the background and for the signal from B.
Once again we would remove a large number of events from this source by excluding the events with other charged tracks around the three muons.

\begin{center}
\begin{figure}[h]
\includegraphics[width=70mm,angle=0]{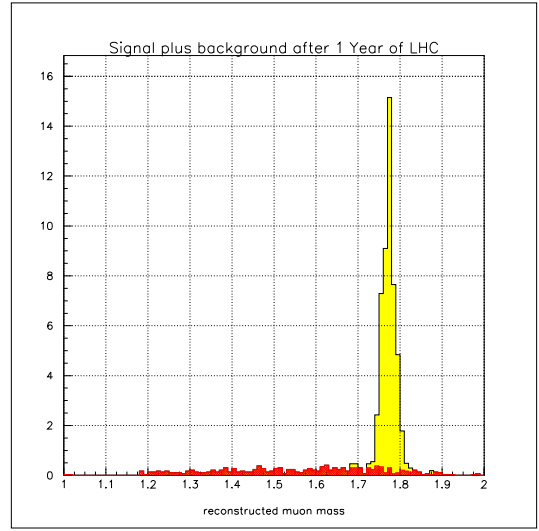}
\caption{Signal plus background after one year of luminosity of LHC assuming a BR equal to the CLEOII limit\label{fig:sigplus}}
\end{figure}
\end{center}

In figures~\ref{fig:chi} and~\ref{fig:massphi} we illustrate the secondary vertex cut and the veto on phi mass.
The probability to have three muons from a common secondary vertex is shown in figure~\ref{fig:chi}. From this figure we also see how signal and cc events present the same distribution while, as expected, the bb events have a peak at zero.
In this case the muons pair originating from Ds decay presents a vertex position distinct from the origin position of the third muon coming directly from Bs decay. 
The dimuon-mass plot for signal and main background are self-explaining and do not need any other comment.
Finally we remove all the background events by applying the three muons mass cut where the resolution found for the signal is 15MeV.
At the end of the analysis, with the hypothesis that the BR of tau in three muons is the actual experimental limit set by CLEO II (1.9$\times 10^{-06}$), we expect after one year 44$\pm 2$ signal events against 1$\pm 1$ of the background.
In fig~\ref{fig:sigplus} we illustrate the plot of the signal plus background and background as expected with an integrated luminosity of 10$fb^{-1}$ (available after one year of LHC running at $\mathcal{L}=10^{33}cm^{-2}s^{-1}$) and  assuming the Branching Ratio to be equal the present CLEOII experimental limit. This will correspond to an upper limit on the BR (at 90\% CL) of 8.4$\times 10^{-8}$.  

\subsubsection{$\tau$ from Z and from B}
In this case we can exploit the second tau which is selected by applying a second isolation criterium based on the fact the tau usually decay in one or three well collimated charged tracks.
Moreover this tau and the tau signal will give the reconstructed Z-mass.
The analysis used for this source is quite similar to that of the W source with the main difference that we do not apply the missing energy cut but we require the following additional cuts:
\begin{itemize}
\item second tau isolation (1 to 3 tracks inside a narrow cone of 0.03 aperture and no other tracks inside a broader and complementary cone of opening 0.4 with Pt$>$1.5
\item Pt$_{\tau}>$23GeV
\item reconstructed mass of the  tau-jet+ missing energy+three muons greater than 70GeV
\end{itemize}

\begin{center}
\begin{figure}[h]
\includegraphics[width=70mm,angle=0]{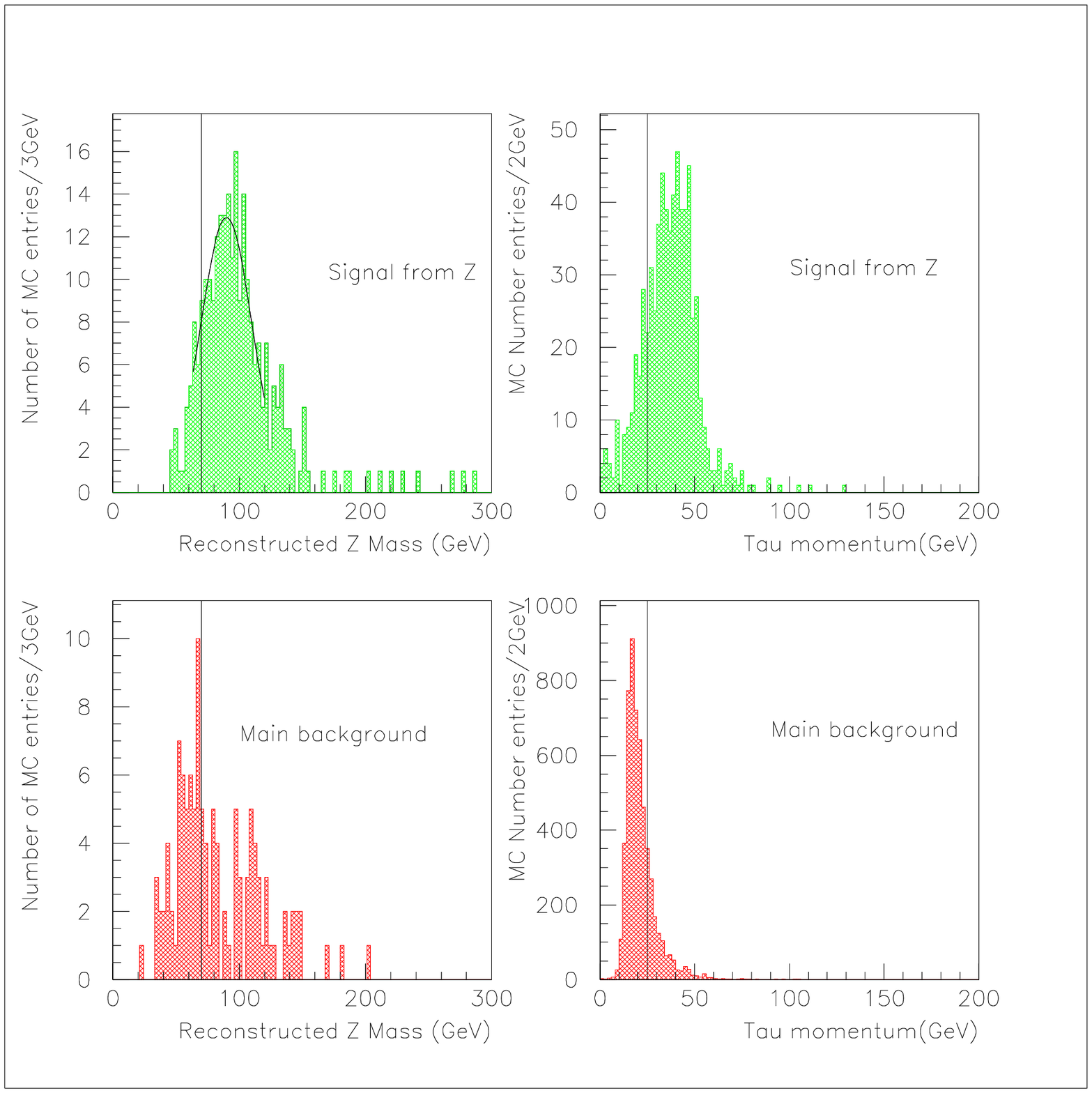}
\caption{Main kinematical variables used for the Z-dedicated analysis \label{fig:zana}}
\end{figure}
\end{center}

In figure~\ref{fig:zana} we have a summary of the Z-dedicated set of cuts. In the upper part we have the distribution for the signal while in the bottom that for main background.

Because of the lower Z cross section production, the final number of events from Z surviving our analysis will be 4$\pm 1$ and 0.8$\pm 1$ the background events for 10$fb^{-1}$. The corresponding sensitivity if no signal is detected will be 7.6$\times 10^{-7}$.

\begin{center}
\begin{figure}[h]
\includegraphics[width=70mm,angle=0]{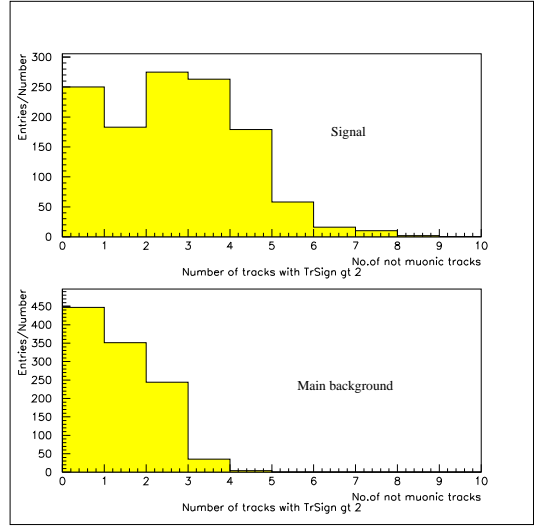}
\caption{Number of tracks with a significance of IP greater than 2 \label{fig:bana}}
\end{figure}
\end{center}

The final sorce of signal considered was the B.
At low luminosity regime of collider this could be a very important source of signal.
These events present a signature completly different from the previous ones.
First, their multiplicity is too high to apply isolation cut and the missing energy selection can not be applied.
Second the energy of the tau is not so high to be a good cut candidate.
In this case after the trigger and the identification criteria, we therefore require two consecutive b-tag for the two b-jet candidate.
The b-tag alghorithm considered was the simplest possible one by requiring some track for every events inside a jet with a significance of the transverse impact parameter greater than a given value to tune.
In figure~\ref{fig:bana} we show this variable for the three muons jet for signal events (top) and main background (bottom). We cut out events presenting less than 3 tracks with significance of impact parameter greater than 2.
With this analysis we can to reach a sensitivity of BR at 90\% equal 3.8$\times 10^{-7}$.
  
\subsection{Instrumental background}
Till now we estimated the sensitivity by considering only physical channels whose experimental signature is closest to that of the signal.
However, tied to very low sensitivity we would reach, it is dangerous to exclude a priori other source of background arising in an experimental environment.
Although,(see \cite{TDR}), the estimation of the probability to mistag at CMS trigger level 1 a light meson (pion or kaon) as muon is very low ($\sim \frac{1}{100}$) and decreases with the threshold on the roughly reconstructed level 1 transverse momentum of the particle, we have done a systematic analysis to evaluate their contribution.
First we generated 500 events, as kinematically close as possible to the signal, and considered the decay  $\tau \rightarrow \pi \pi \pi \nu_{\tau}$ (with the tau from W) replacing two pions with two real muons.
In this way we are able to estimate the probability to mistag the third $\pi$ as $\mu$ by simply counting how many close muons have been reconstructed by our simulation program at level 1 or 2 or 3.
In figure 9 we show these numbers.\\
\begin{center}
\begin{figure}[h]
\includegraphics[width=70mm,angle=0]{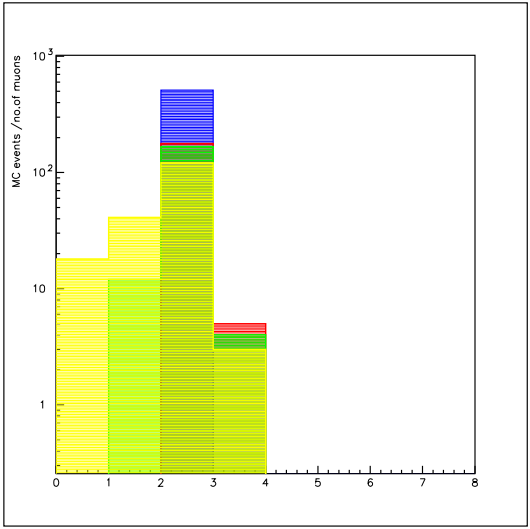}
\caption{Number of real, L1, L2 and L3 reconstructed muons  \label{fig:fake}}
\end{figure}
\end{center}
We found at level-1 that 5 events (among 500) present 3 $\mu$. We can then estimated a 1\% misedintification probability.\\
After this cross-check, we have considered as background the following channel, $\tau \rightarrow \pi \pi \pi \nu_{\tau}$, with all the three pions misidentified as muons.
The expected number of events passing after one year the analysis (W dedicated) is given by the formula:
\begin{eqnarray}
N_{1year}=\mathcal{L}\times BR(pp\rightarrow W\rightarrow \tau \nu)\times  \nonumber \\BR(\tau \rightarrow \pi \pi \pi \nu)\times 
P(\pi \rightarrow \mu)^3 \times \epsilon_{analysis}
\end{eqnarray}

With $\epsilon_{analysis} \simeq 10^{-3}$ and BR$\sim 9\%$, we found for these events an expected rate negligible compared with the physical background.
\section{$\tau \rightarrow \gamma \mu^- $}
Encouraged by the previous results we have undertaken a preliminary study of the possibility to detect the $\tau \rightarrow \mu \gamma$ decay channel, which is expected to be much higher even if much harder to be detected experimentally. We have focused on the $\tau$ source which allows to exploit the large missing energy expected.
So the background has to be seeked inside muons sources with high missing energy.
The same background should have a very hard and well isolated muon.
From these preliminary considerations we focused our attention to the following events:
\begin{itemize}
\item $pp \rightarrow W \rightarrow \mu \nu_{\mu}$  (SAMPLE A $\sigma =19nb$)
\item $pp \rightarrow W \gamma \rightarrow \mu \gamma \nu_{\mu}$ (SAMPLE B $\sigma=0.18nb$)
\item $pp \rightarrow W \rightarrow \tau \nu_{\tau} \rightarrow \mu \nu_{\mu} \nu_{\tau} \nu_{\tau}$ (SAMPLE C $\sigma =2.1nb$)
\end{itemize}

These events present similar kinematical features although some important differences arise.
The sample A present the highest cross section. The photon presence is not assured and eventually it could come from some $\pi^0$.
The sample B, with a photon radiated from primary interaction, present a cross section 100 times smaller.
The muon is very hard and the photon is not always close to muon.
The sample C is expected to have the most signal like distributions and then the rejection power to be the poorest among the channels considered.
The analysis steps are the following
\begin{itemize}
\item Level 1 trigger
\begin{enumerate}
\item 1 photon with L1 E$_T>25$ GeV or
\item 1 muon with L1 p$_T>20$GeV or
\item 1 muon with L1 p$_T>5$GeV and 1 photon with L1 E$_T>15$ GeV
\end{enumerate}
\item high level trigger:1 well reconstructed photon (level 2) and one muon L3 or just one high transverse momentum track close to the photon
\item reconstructed transverse energy of photon greater than 18 GeV
\item missing transverse energy greater than 20 GeV
\item candidate muon momentum less than 20 GeV
\item angular distance between muon and photon greater than 008 and less than 0.15
\item significance of impact parameter of muon greater than 2
\item reconstructed mu-gamma mass equal the tau mass $\pm$ 60MeV
\end{itemize}
In figure from 10 to 12 we illustrate the distributions of the quantities used in the analysis.
The sample C of background is shown in dashed line in figures 10 and 11.
In solid line we plotted the same variables for sample A events and in dotted line for sample B.
We note that the sample C shows distributions for the $\mu$ momentum and transverse impact parameter closest to these for the signal.
In figure 12 and 13 we plotted the transverse energy of photon for the
signal (filled area) and for the background (solid line) and angular distance between muon and photon.
We avoided to show for these quantities the distribution for sample B
and C whose shape is similar to the sample A.
\begin{center}
\begin{figure}[h]
\includegraphics[width=70mm,angle=0]{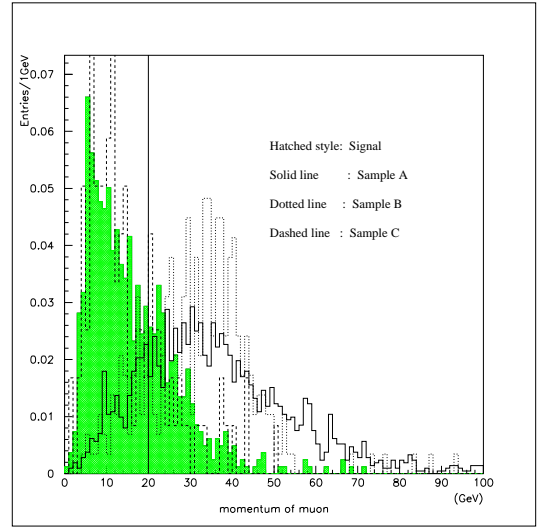}
\caption{Muon momentum distribution \label{fig:momentum}}
\end{figure}
\end{center}
\begin{center}
\begin{figure}[h]
\includegraphics[width=70mm,angle=0]{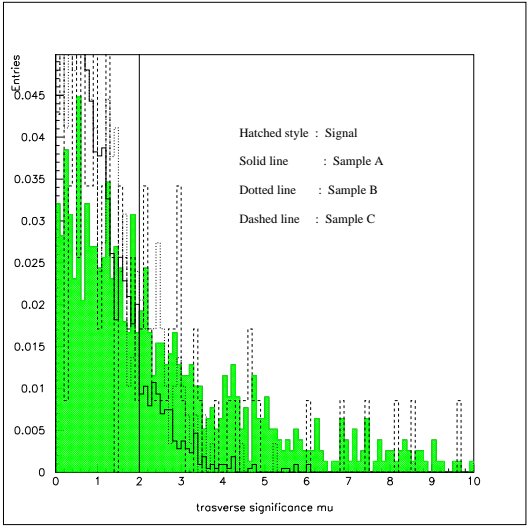}
\caption{Muon transverse impact parameter significance distribution \label{fig:trip}}
\end{figure}
\end{center}

\begin{center}
\begin{figure}[h]
\includegraphics[width=70mm,angle=0]{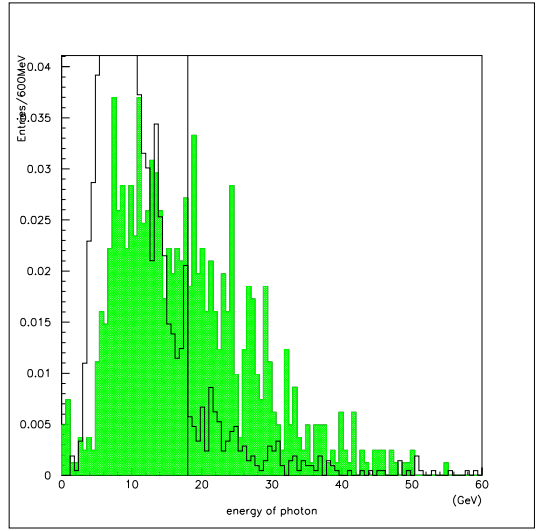}
\caption{Photon transverse impact energy distribution (filled area for the signal)\label{fig:energy}}
\end{figure}
\end{center}
\begin{center}
\begin{figure}[h]
\includegraphics[width=70mm,angle=0]{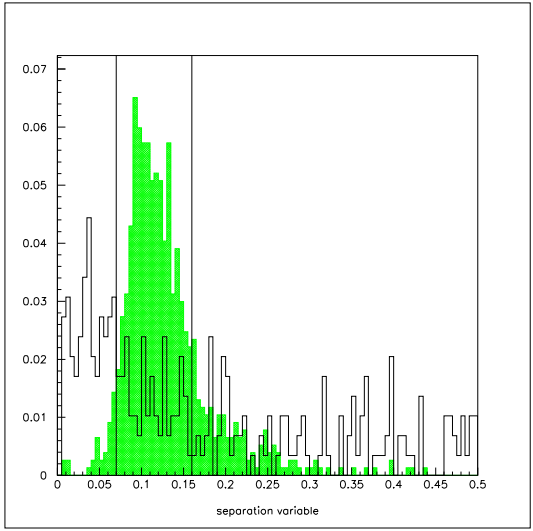}
\caption{Photon-muon angular distance distribution (filled area for the signal)\label{fig:distance}}
\end{figure}
\end{center}
\begin{center}
\begin{figure}[h]
\includegraphics[width=70mm,angle=0]{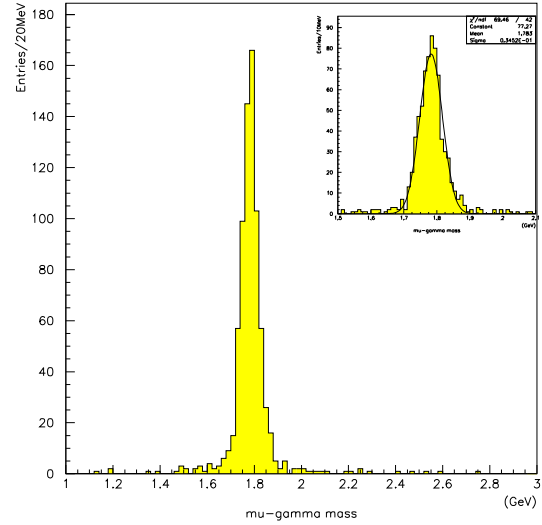}
\caption{Signal muon-photon reconstructed mass\label{fig:taumas1}}
\end{figure}
\end{center}
\begin{center}
\begin{figure}[t]
\includegraphics[width=70mm,angle=0]{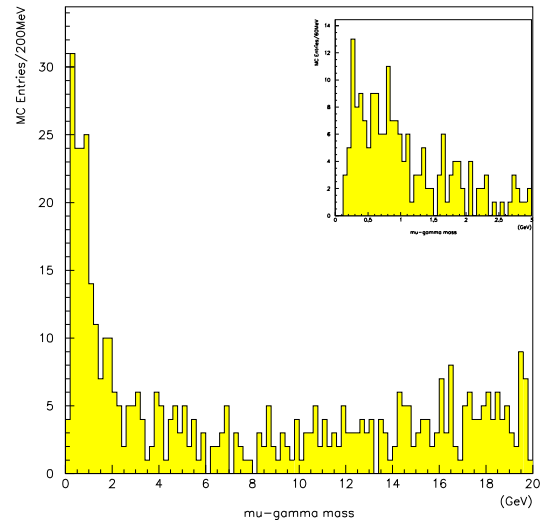}
\caption{Background muon-photon reconstructed mass\label{fig:taumas2}}
\end{figure}
\end{center}

In figure 14 we present the photon-muon reconstructed mass for the
signal events. We also plot the mass region around the tau mass which
shows a resolution of around 40 MeV.
(this is worse than one found in CLEOII and Atlas collaboration \cite{Atlas}) 
At the end of the analysis we still remain with around a 6\% of 
MonteCarlo initial signal events produced and we remove all the initial 40000 events of sample A
and 3500 events of sample B and C.
The normalized number of background events expected after one year
is 13 from sample A, less than 1 from B and 18 events from sample C
which is confirmed to be the most dangerous background.
To find these numbers we adopted the hypothesis the mass distribution
for the background is uniform in the range between 0 to 3 GeV.
This is a rather pessimistic hypothesis,
because the mass for the background (upper box in fig.~\ref{fig:taumas2}) is peaked below 1 GeV/$c^2$ and in the interesting region ($\tau$-mass $\pm$ 60MeV/$c^2$) the number of events is smaller than the one foreseen with a flat distribution assumed. 
With these assumptions we evaluated the BR to be less than $10^{-6}$ for the $\tau \rightarrow \mu \gamma$ after one year of low luminosity of LHC taking data.(10 fb$^{-1}$ )\\
\section{Conclusion}

We have presented the MonteCarlo based analysis of neutrinoless decays channel $\tau \rightarrow \mu^+ \mu^- \mu^-$ and  $\tau \rightarrow \mu \gamma$ using the CMS object-oriented reconstruction program ORCA.
We isolated and analysed three important sources of signal.
We found that the most dangerous background for the first channel is represented by $c \overline c$ events with three muons coming from a common meson. After one year of low luminosity of LHC we could improve by a factor 25 the actual experimental sensitivity to this channel by using only the W source.
Other channels (Z and B) give a minor contribution in the present analysis.
For the $\tau \rightarrow \mu \gamma$ channel the most dangerous background arises from leptonic decays of W boson and a first study confirm the  results of ATLAS.
The signal is background limitated and the CMS sensitivity for this analysis will be similar to that expected at the B factories.
Nevertheless, more detailed studies of pile-up effects are required to extend the analysis to higher luminosity.


\begin{thebibliography}{9}
\bibitem{Ellis}  hep/ph 9911459 J.Ellis at al.
\bibitem{nuclear}  Nucl.Physics B345 331-368
\bibitem{muons}  CMS-NOTE 1997/096
\bibitem{TDR} CERN/LHCC 2000-38 CMS TDR 6.1
\bibitem{Atlas} Atlas internal Note PHYS-97-114
\bibitem{PYTHIA} T.Sjostrand, CERN-TH/7112/93
\bibitem{CMSIM} http://cmsdoc.cern.ch/cmsim/cmsim.html
\bibitem{GEANT} Cern Program Library Writeup W5013
\bibitem{ORCA} http://cmsdoc.cern.ch/orca
\end{thebibliography}
\end{document}